\documentclass[11pt]{revtex4}
\usepackage{graphicx}
\usepackage{amsmath}
\usepackage{amssymb}

\begin{document}

\title{Using Abrupt Changes in Magnetic Susceptibility within Type-II Superconductors to Explore Global Decoherence Phenomena}

\author{\large{Stephen J. Minter$^a$, Raymond Y. Chiao$^b$, and Luis A. Martinez$^c$} \\
\vspace{0.1in}
\scriptsize{$^a$Vienna Center for Quantum Science and Technology, Faculty of Physics, \\University of Vienna,
Boltzmanngasse 5, A-1090 Vienna, Austria\\
Corresponding author, email: stephen.minter@univie.ac.at \\
\vspace{0.05in}
$^b$University of California, Merced, Schools of Natural Sciences and Engineering, P.O. Box 2039, Merced, CA  95344, USA\\
\vspace{0.05in}
$^c$University of California, Merced, School of Natural Sciences, P.O. Box 2039, Merced, CA  95344, USA\\}
\vspace{0.1in}\normalsize{PACS: 74.90.+n, 03.65.-w, 03.30.+p}}

\begin{abstract}\textbf{Abstract}: A phenomenon of a periodic ÒstaircaseÓ of macroscopic jumps in the admitted magnetic field has been observed, as the magnitude of an externally applied magnetic field is smoothly increased
or decreased upon a superconducting (SC) loop of type II niobium-titanium wire which is coated
with a non-superconducting layer of copper. Large temperature spikes were observed to occur
simultaneously with the jumps, suggesting brief transitions to the normal state, caused by en masse motions of Abrikosov vortices. An experiment that exploits this phenomenon to explore the global decoherence of a large superconducting system
will be discussed, and preliminary data will be presented.  Though further experimentation is required to determine the actual decoherence rate across the superconducting system, multiple classical processes are ruled out, suggesting that jumps in magnetic flux are fully quantum mechanical processes which may correspond to large group velocities within the global Cooper pair wavefunction.
\end{abstract}

\maketitle




\section{Instability Regions in Type-II Superconductors}

When the magnitude of an externally applied magnetic field incident on a closed loop of a Type-II superconductor (such that the field has a non-zero component parallel to the axis of the loop) exceeds the first critical field $H_{\text{c1}}$, Abrikosov vortices of Cooper pair electrons will form, allowing some flux lines to penetrate the sample.  When the field is increased with time, the motion of the Abrikosov vortices can give rise to a slight increase in thermal energy and therefore a slight increase in temperature.  Since the London penetration depth

\begin{equation}
\lambda_L(T) = \lambda_L(0)[1-(T/T_c)^4]^{-1/2}
\end{equation}
where $T$ is the sample temperature, and $T_\text{c}$ is the superconducting critical temperature, increases with the positive change in temperature, lines of magnetic flux from the applied field push further into the sample.  In addition, the critical current density

\begin{equation}
j_\text{c}(T)=j_\text{c}(0)(1-T/T_\text{c})
\end{equation}
decreases, an effect which creates an electric field via the London equation 

\begin{equation}
{\bf E}=\frac{\partial}{\partial t}(\Lambda{\bf j})
\end{equation}
This electric field can lead to another temperature increase, and the process will repeat in a cascading effect until the magnetic flux line densities in the two regions are such that

\begin{equation}
B_{\text{out}}=B_{\text{in}}+\mu_0H_{c1}
\label{equilibrium}
\end{equation}
where $B_{\text{out}}$ is the field magnitude just outside of the loop, and $B_{\text{in}}$ is the field magnitude inside the loop.  This process, which restores the condition given in (\ref{equilibrium}), is henceforth referred to as a \textquotedblleft flux jump\textquotedblright .  The increase in temperature that accompanies a flux jump drives the sample, or part of the sample, to a normal state in most cases \cite{Mints1}, resulting in the collapse of the global Cooper pair wavefunction state that spanned the entire superconducting electron gas before the flux jump occurred.  While the flux jumps are thermal effects that are not discontinuous, the rate of increase of the magnetic induction inside the loop during a flux jump can be assumed throughout this article to satisfy

\begin{equation}
\dot{B}_{\text{in}}(t)\gg \dot{B}_{\text{a}}(t)
\end{equation}
where $\dot{B}_{\text{a}}$ is the ramping rate of the externally-applied field.

Due to the phenomenon of \textquotedblleft flux creep\textquotedblright, in which lines of magnetic flux pass, via Abrikosov vortices, from the outside of the loop, where the field is higher, to the inside of the loop, but do not raise the temperature so as to cause a flux jump, condition (\ref{equilibrium}) can be satisfied without the occurrence of flux jumps for small ramping rates $\dot{B}_{\text{a}}$.  There exists a maximum ramping rate for which the adiabatic flux creep process is stable, below which no flux jumps will occur, which is given by \cite{Mints2}\cite{adiabatic-note}

\begin{equation}
\dot{B}_{\text{a}}<\frac{8}{\pi^2}\frac{\mu_0j_1h}{C}
\label{adiabatic-condition}
\end{equation}
where $\mu_0$ is the permeability of free space, $j_1=E\frac{\partial j}{\partial E}$, where $E$ is the electric field induced in the flux creep process, $h$ is the heat transfer coefficient, and $C$ is the heat capacity of the sample.

While it can be assumed that flux jumps will occur when the condition in (\ref{adiabatic-condition}) is not met (by using larger ramping rates that are above this threshold), and this has been reproducibly verified by our experiments, an individual flux jump is a quantum-mechanical process, and the time at which one will occur cannot be reliably predicted \emph{a priori}.  This is due to the fact that a flux jump represents a change between quantum states (via persistent current modes), much like spontaneous emission in a two-level atom.

While the onset of a flux jump cannot be predicted ahead of time to arbitrary accuracy, one can detect the onset of a flux jump after it has occurred by measuring the abrupt change in magnetic field created by the change in persistent currents.  For example, prior to the first flux jump, a persistent current will flow around the outside edge of a superconducting ring in order to preserve the absence of magnetic flux lines within the closed loop (ignoring for the moment the flux creep process).  As the field continues to increase, so will the persistent current.  When a flux jump occurs, a persistent current will be established on the \emph{inside} edge of the ring to, again, preserve the number of flux lines that have entered the closed loop, until the next flux jump occurs.  The sudden increase in the persistent current along the inside edge, coupled with the sudden decrease in the persistent current along the outside edge, allows for straightforward measurement of the flux jump process via a magnetic field sensor or pickup coil that is placed coaxially with the closed loop.   Modeling the loop as a pure magnetic dipole, the change in field at a distance $d$ from the loop is given by

\begin{equation}
\Delta B=\frac{\mu_0r^2}{4d^2}\Delta I
\end{equation}
where $r$ is the radius of the loop.  $\Delta I$, which should take into account the changes in persistent currents on both the inside and outside edges of the superconducting loop, will depend on the ramping rate $\dot{B}_{\text{a}}$ and the critical fields of the superconductor.  A detailed, quantitative analysis can be found in \cite{Mints1}.  Since the rate of increase of the admitted magnetic flux line density $\dot{B}_{\text{in}}$ tends to be large, the back-emf in a pickup coil, in accordance with Faraday's law, is typically straightforward to detect with an oscilloscope.

An interesting situation arises when one considers a closed loop of superconducting wire that experiences a large change in magnetic field $\dot{B}_{\text{a}}$ upon only a part of the system.  In this particular experiment, a superconducting wire was formed such that one coil was wound, then a second coil wound with 3 meters of wire in between the two coils, and the wire routed back to the first coil and joined to the starting point to form a complete, coherent superconducting loop.

One coil (henceforth referred to as the \textquotedblleft low field SC coil\textquotedblright ) was placed in a region where the ramping rate $\dot{B}_{\text{a}}$ was small, namely, near the null of a quadrupolar anti-Helmholtz field, while the other (the \textquotedblleft high field SC coil\textquotedblright ) was placed in a region of high ramping rate, below the lower magnet in the anti-Helmholtz pair, where the axial field is maximal.  Thus, the high field coil would experience flux jumps for certain ramping rates while the low field coil would not if each were a \emph{separate}, closed coil.  However, this is not the case, since the two-coil system is made from a single, coherent superconducting loop.  Therefore, any changes in persistent currents that arise from the flux jumps induced in the high field coil, as well as any collapse of the global Cooper pair wavefunction, must also affect the low field coil.

Since the Cooper pair wavefunction must be single-valued everywhere along a continuously-connected superconducting system \cite{byers}, a collapse of the wavefunction is a global event, even if it is triggered locally, e.g. by a flux jump that drives a section of the wire above the superconducting critical temperature.  The question arises, can one determine how quickly the wavefunction collapse occurs in the low field coil if it is triggered in the high field coil?  Furthermore, can one determine how quickly the persistent currents disappear when the coherent superconducting connection is broken by a flux jump within the high field coil?  The answers to these questions involve considerations of relativity, since a global collapse of the wavefunction triggered by a local event implies instantaneous action-at-a-distance.

\section{Large Group Velocities Within Superconductors}

In 1905, Einstein published his historic paper on special relativity.
Shortly afterwards, Sommerfeld \cite{Brillouin}\ answered criticisms of
Einstein's work, namely, that the phase and group velocities of
electromagnetic waves can become superluminal, since these two kinds of
velocities can exceed the vacuum speed of light\ inside a dielectric medium.
Note that Einstein considered wave propagation solely in the vacuum,
whereas his critics considered wave propagation in media.

Sommerfeld pointed out that while it is true that both the phase and the
group velocities in media can in fact exceed $c$, the \emph{front} velocity,
defined as the velocity of a \emph{discontinuous jump} in the initial wave
amplitude from zero to a finite value, cannot exceed $c$. It is Sommerfeld's
principle of the non-superluminality of the front velocity
that prevents a violation of the Einstein's basic principle of causality in
special relativity, i.e., that no effect can ever precede its cause. (See 
\cite{Chiao tutorial}).

In subsequent work, Sommerfeld and Brillouin \cite{Brillouin} showed that
the \textquotedblleft front\textquotedblright\ is accompanied by two kinds
of \textquotedblleft precursors\textquotedblright , now known as the
\textquotedblleft Sommerfeld\textquotedblright , or the \textquotedblleft
high-frequency\textquotedblright , precursor, and the \textquotedblleft
Brillouin\textquotedblright , or the \textquotedblleft
low-frequency\textquotedblright , precursor. These precursors are weak
ringing waveforms that follow the abrupt onset of the front, but
although they can precede the gradual onset of the strong main
signal, they can never precede the onset of the front. One can
therefore view the precursor phenomenon as a kind of \textquotedblleft
shock-wave\textquotedblright\ response to the collision between the analytic
portions of the waveform with the nonanalytic, discontinuous front of the
waveform. Such \textquotedblleft shock-wave\textquotedblright\ waveforms,
however, can never pass through the front and somehow come ahead of the
front. This then is meaning of Einstein causality.

It is well known that the phase velocity of electromagnetic waves can become
superluminal under certain circumstances. A simple example is the
superluminality of the phase velocity of an electromagnetic wave traveling
within a rectangular waveguide in its fundamental TE$_{01}$ mode. Another,
more impressive, example is the superluminality of the phase velocity of
X-rays in all materials. As was first noticed by Einstein, the
superluminality of the phase velocity of X-rays in all kinds of crystals
leads to the phenomenon of total external reflection of X-rays
impinging at grazing incidence from the vacuum upon the surface of any kind
of crystal \cite{Chiao tutorial}. Hence the superluminality of the phase
velocity has physically observable consequences.

However, one cannot send a true signal faster than light by means of a
superluminal phase velocity, since the phase velocity is the velocity of the
crests (i.e., the phasefronts) of a continuous-wave, monochromatic,
electromagnetic wave. Since the amplitude and phase of a continuous wave do
not change with time, there can be no information contained within such an
infinite waveform. As in radio, one must introduce a time-dependent modulation of the continuous \textquotedblleft
carrier\textquotedblright\ waveform (using either AM or FM modulation),
i.e., a genuine \emph{change} in the carrier waveform, before any true
signal can be sent via the carrier wave.

While it is well known that phase velocities can become superluminal, it is
less well known that group velocities can also become superluminal. There is
a common misconception that the group velocity is the \textquotedblleft
signal\textquotedblright\ velocity of physics, which relates a cause to its
effect, and therefore that it cannot propagate faster than light. However,
the group velocity is not the velocity that relates a cause to its
effect. Only Sommerfeld's front velocity can fulfill this role.

One experimentally observed example of the occurrence of superluminal group
velocities is that individual photons tunnel superluminally through a tunnel
barrier \cite{Steinberg}. There have been numerous other
observations of superluminal group velocities of laser pulses propagating
superluminally and transparently through various kinds of dielectric media
with optical gain \cite{Boyd and Gauthier}. A recent example of superluminal
group velocities is the observation of the superluminal and transparent
propagation of laser pulses within optical fibers which possess stimulated
Brillouin gain \cite{Shanghai superluminality}.

Nevertheless, Sommerfeld showed that it is the \emph{front} velocity, and
only the front velocity, that relates a cause to its effect in special
relativity. He introduced the theta function%
\begin{equation}
\Theta (t)=\left\{ 
\begin{array}{c}
0\text{ for all times }t<0 \\ 
1\text{ for all times }t\geq 0%
\end{array}%
\right. 
\end{equation}%
in order to modulate any kind of continuous carrier wave. The instant $t=0$
corresponds to the sudden turn-on of the carrier wave, initiated, for
example, by the pushing of the \textquotedblleft ON\textquotedblright\
button of a continuous-wave radio-frequency signal generator. This
\textquotedblleft push-button\textquotedblright\ kind of signaling guarantees that no effect can precede its cause. The light-cone
structure of spacetime in relativity follows from the propagation of these
\textquotedblleft push-button signals\textquotedblright\ at the front
velocity, and not from the propagation of smooth, analytic \textquotedblleft
wavepacket signals\textquotedblright\ at the group velocity, such as the
superluminal propagation of a Gaussian wavepacket within a transparent
dielectric medium with gain in it. Hence the \textquotedblleft
signal\textquotedblright\ velocity of physics, in the fundamental sense of a
physical \textquotedblleft signal\textquotedblright\ that connects a cause
to its effect, is given by the front velocity, and not by the group
velocity.

When a flux jump occurs, the Cooper pair wavefunction collapses across the entire superconducting system, and thus a global change in phase accompanies a flux jump.  However, note that one is \emph{not} directly measuring the collapse of the wavefunction, but the change in supercurrent density (in this particular experiment, via a back-emf voltage induced in a pickup coil due to the changing magnetic flux).  Thus, it is not the change in global phase $\nabla \phi$ which is an observable, but a change in the value of the current density ${\bf j}$.  The two are related via the minimal coupling rule by
\begin{equation}
{\bf j} = \frac{\rho}{m^*}\left(\frac{\hbar}{i}\nabla \phi - q^*{\bf A}\right)
\label{min-coupling-rule}
\end{equation}
where the general form of the Cooper pair wavefunction is $\psi=\sqrt{\rho}e^{i\phi}$, ${\bf A}$ is the magnetic vector potential, and $m^*$ and $q^*$ are the mass and charge of a Cooper pair, respectively.  Thus, while the two terms on the right side of (\ref{min-coupling-rule}) may undergo changes via an instantaneous action-at-a-distance process, the observable {\bf j} remains unchanged until after the luminal or sub-luminal \textquotedblleft wavefront\textquotedblright (the signal that propagates from the high field SC coil to the low field SC coil immediately following the onset of a flux jump, which travels with velocity $v \le c$).

\begin{figure}[tbh]
\centering
\includegraphics[angle=0,width=.4\textwidth]{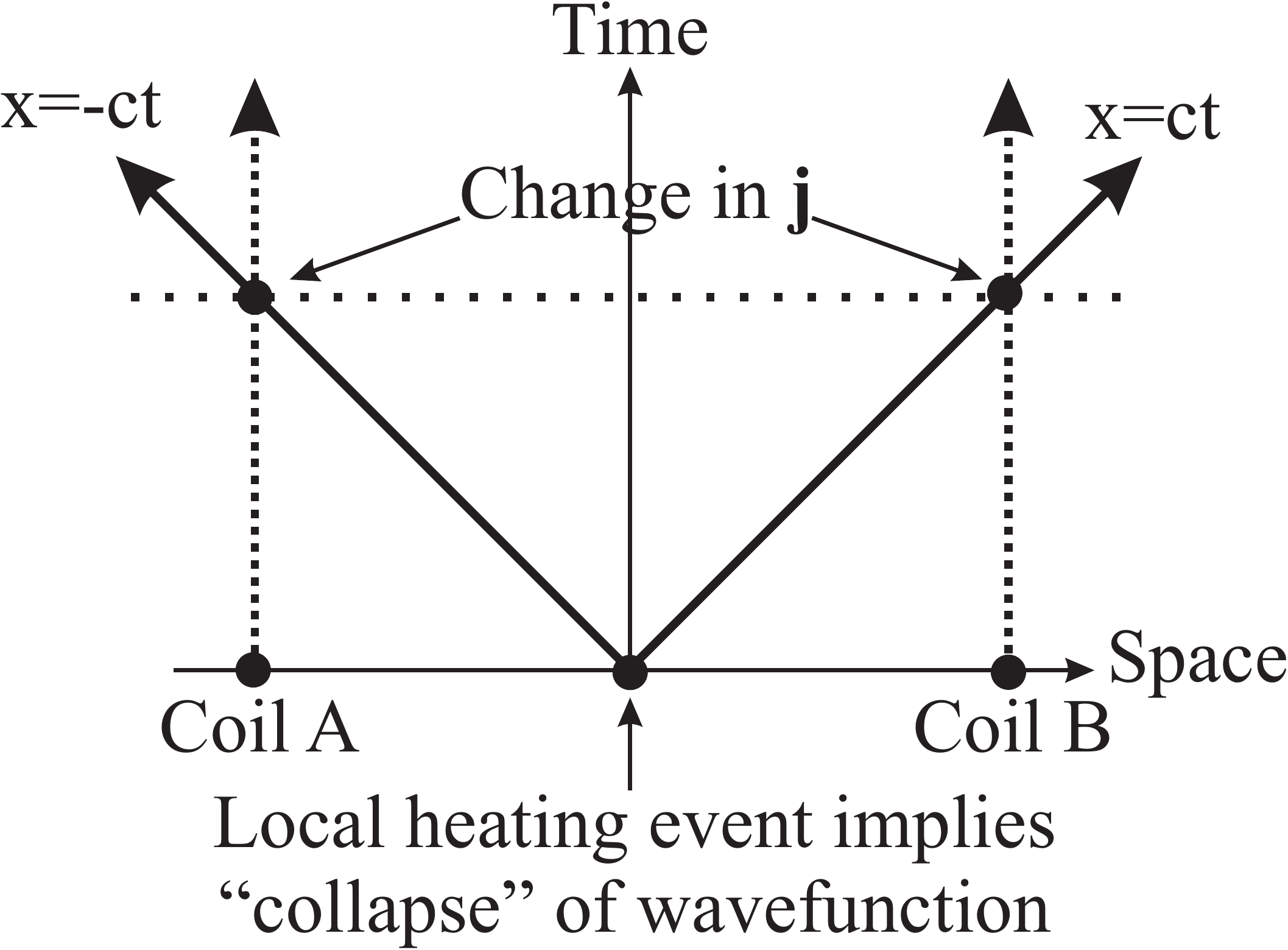}
\caption{A local heating event that leads to
a \textquotedblleft collapse\textquotedblright\ of the wavefunction
everywhere within a coherent SC circuit connecting SC coils A and B, cannot
instantly cause the persistent current $\mathbf{j}$ to disappear everywhere
instantaneously, even if changes in the vector potential and the phase were to occur outside of the light cone.}
\label{fig:spacetime}
\end{figure}

A simplified picture is shown in Figure \ref{fig:spacetime} where, for simplicity and without loss of generality, the decoherence event occurs between the two coils which have been relabeled \textquotedblleft Coil A\textquotedblright and \textquotedblleft Coil B\textquotedblright to establish symmetry.  While the decoherence event may create a change in global phase and a change in the electromagnetic vector potential at the two coils on the space axis, no faster-than-light signals can be sent since these quantities cannot be freely manipulated at the origin nor measured at the coils.  A change in the current density {\bf j} can be detected at either coil, but the measurement must take place inside the light cone of the initial decoherence event at the origin of the spacetime diagram if it is to be observed.

One might wonder whether the instantaneous change in the vector potential could be measured via the Aharonov-Bohm effect in which, for example, an electron acquires a phase in the presence of a non-zero vector potential, but in the absence of a local magnetic field.  However, recall that the Aharonov-Bohm experiment allows one to measure the line integral $\oint{\bf A}\cdot d{\bf l}$ which is equal to the magnetic flux (obvious after an application of Stokes' Theorem), and \emph{not} the vector potential ${\bf A}$ itself.  Thus, the phase incurred through the Aharonov-Bohm effect would not be measurable until the flux changes, which is due to a change in {\bf j}, which, as discussed, can only occur on a time scale less than or equal to $l/c$, where $l$ is the distance between the coils, again forbidding any faster-than-light signaling.



%

\section{Appendix: Preliminary Experimental Data}

A preliminary experiment was conducted in which the abrupt change in magnetic field $\dot{B}_{\text{in}}$ was measured (via voltage signals in nearby pickup coils) within both the high field and low field SC coils during a flux jump.  Attempts were made to measure the difference in time between the leading edges of each signal to determine the delay in arrival between the two voltage signatures.  For specific experimental parameters, see \cite{data}.

\begin{figure}[tbh]\centering\includegraphics[width=.5\textwidth]{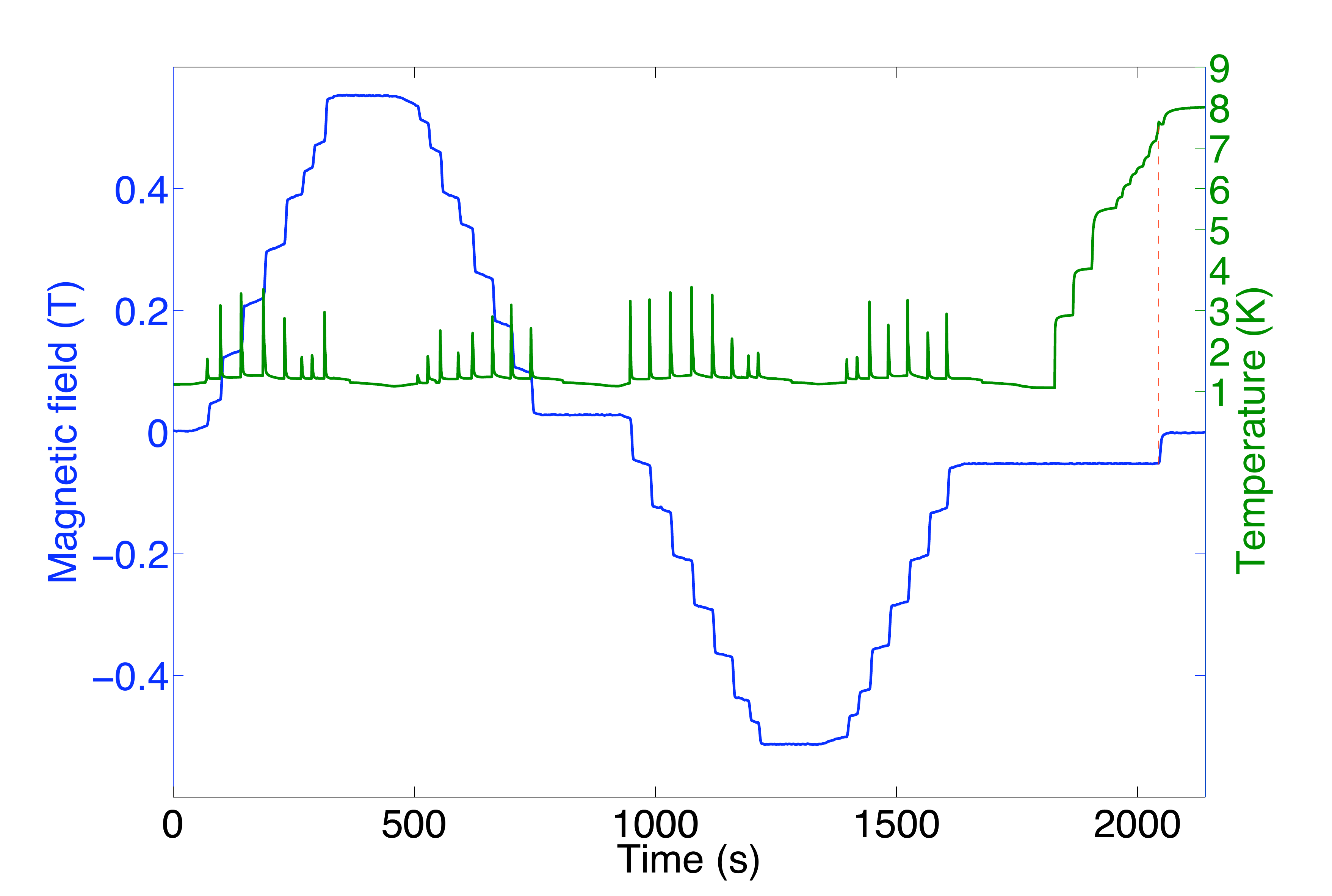}\caption{Simultaneous plots of measured magnetic field $B_{\text{in}}$ vs. time and temperature vs. time for a single closed coil (of the same SC wire that was used in the actual experiment) during phases of constant ramping rate $\dot{B}_{\text{a}}=0,\pm 390 \text{ G/min}$ (not depicted) in a homogeneous field parallel to the coil axis.  Note that each flux jump corresponds to a temperature increase (whose magnitude appears, unsurprisingly, to be correlated with the magnitude of the flux jump), though this data cannot be used to determine the maximum temperature in the wire during the flux jump due to the low temporal resolution of the temperature measurement (1 sample per second).  Also note that at a temperature of $T\approx 7.5 \text{ K}$, the magnetic field generated by persistent currents due to hysteresis was destroyed.}\label{fig:figL0}\end{figure}

Several voltage signatures were recorded with the pickup coils at different time scales.  The voltage signals were highly reproducible and multiple data sets were averaged to reduce noise levels. Each voltage signal was observed to coincide with an increase in local temperature, as measured by temperature sensors placed on each SC coil. The flux jump phenomena along with the accompanying temperature increases can be seen experimentally in Figure \ref{fig:figL0}. 

\begin{figure}[h!]\centering\includegraphics[width=.5\textwidth]{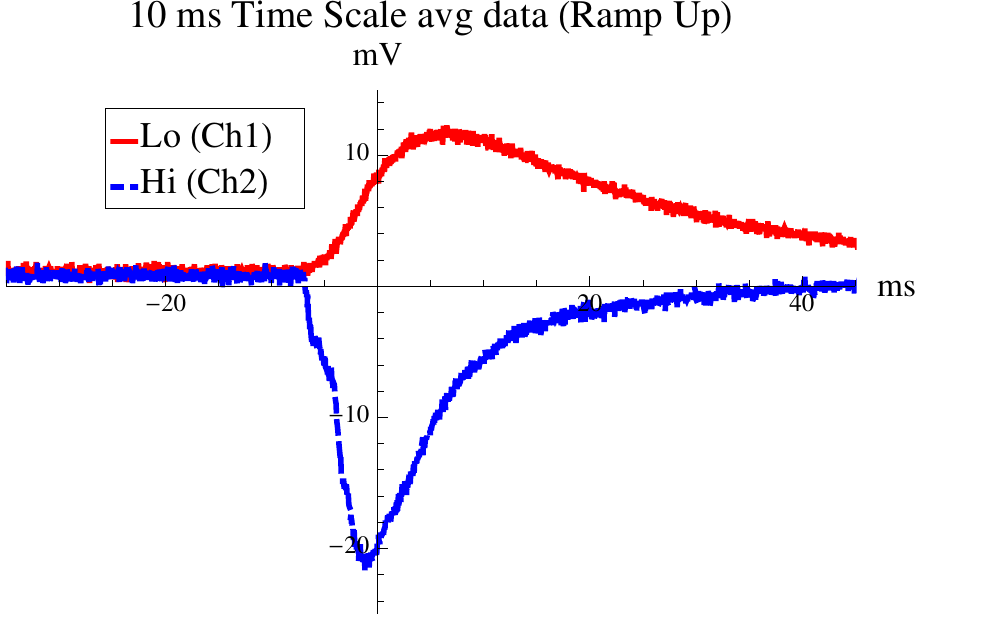}\caption{Average signals at pickup coils for the high and low field SC coils. This particular time scale shows the entire signal during a flux jump during a magnetic field ramp up, and is thus only reliable for characterization of the voltage signatures, and not for determining the time gap between leading edges of each signal. The decay time constant for the high field SC coil (Ch2) is on the order of 10 ms, and 20 ms for the low field coil.}\label{fig:figL1}\end{figure}

Figure \ref{fig:figL1} shows a typical, complete voltage signal as measured by the pickup coil connected to the oscilloscope during a ramp-up of the magnetic field (i.e. $\dot{B}_{\text{a}} > 0)$. The reason for the different polarity in the voltage signals is that the SC coils were oppositely-oriented in order to rule out any false signals due to direct detection of the flux jump in the high field SC coil by the pickup coil at the low field SC coil.  The flux jump causes a temperature increase in the high field SC coil that, in turn, causes the persistent currents to flow into the copper cladding instead of through the superconducting channel, where there are ohmic losses.  However, it is assumed that, over the short time scales in which the flux jump occurs, the currents continue to flow through the superconducting channel in the low field SC coil, where there are no ohmic losses.  Thus, the current decay rate is expected to be larger in the high field coil, which is evident in Figure \ref{fig:figL1}.  At this temporal resolution it is not possible to resolve any time differences between the leading edges of the two signals. 



\begin{figure}[h!]\centering\includegraphics[width=.5\textwidth]{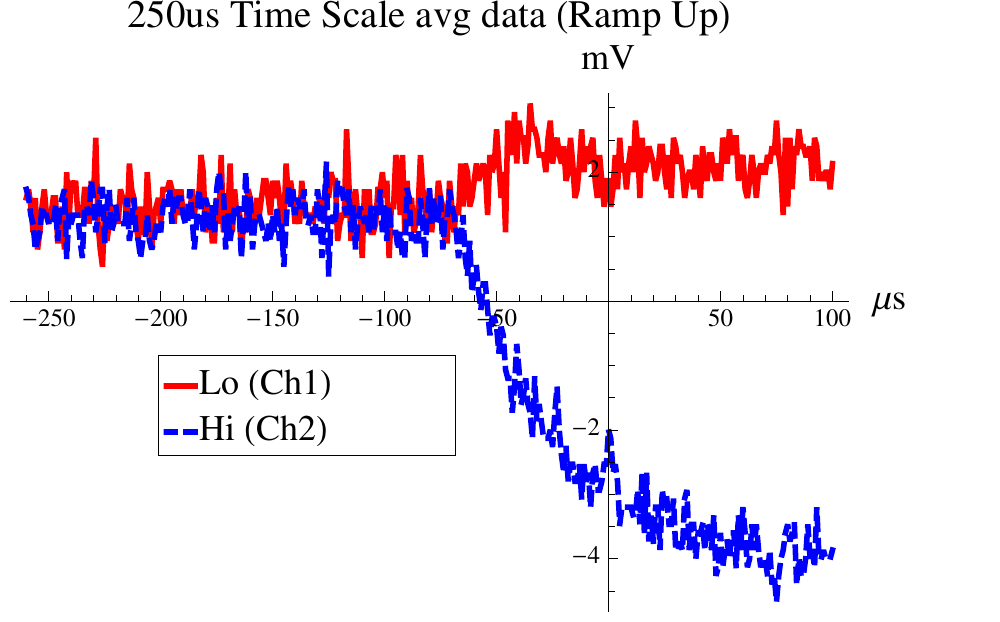}\caption{A shorter time scale than that in Figure \ref{fig:figL1} shows the initial rise of the signals during a magnetic field ramp up. This plot is an average of 3 curves. }\label{fig:figL3}\end{figure}

Figure \ref{fig:figL3} shows the result of 3 averages at a smaller time scale during magnetic field ramp-up. While a highly precise measurement of the delay between the leading edges of the two signals is difficult to determine because of the noise floor, it can be inferred that the difference is on the order of not more than tens of micro-seconds.

There are three other processes which can potentially lead to a voltage signal in the pickup coil near the low field SC coil, instead of the quantum mechanical collapse of the Cooper pair wavefunction, as described in Section II.  However, we will calculate the time scales involved in each case, and show that the actual time delay between the leading edges of the two voltage signals is much faster, suggesting the previously described model is a more likely scenario.  

The first is the time scale related to the inductance-to-resistance ($L/R$) ratio, over which currents would decay in a purely classical circuit.  This is a realistic model for the currents during a flux jump in the high field SC coil, since the currents travel through the copper cladding.  The inductance of the (individual) SC coils is approximately 440 $\mu$H, and the resistance is estimated (and verified through a four-lead measurement) to be approximately 2 m$\Omega$ using the geometry of the wire and the temperature-dependent value of electrical resistivity for copper.  Though a voltage signal can theoretically be seen due to the change in current magnitude before one characteristic time constant has passed, one can see immediately that this time constant is on the order of hundreds of ms, which is \emph{many} orders of magnitude larger than the actual measured time decay, and therefore it is not likely that this is the mechanism under which the voltage signal was generated.

The second alternative considered is phonon interaction in the copper cladding when a flux jump takes place.  Since the collapse of the Cooper pair wavefunction is stimulated in the high field SC coil, the time for a phonon to travel from the high field SC coil to the low field SC coil bears consideration.  This time is simply characterized by $l/v_{\text{s}}$, where $l=3$ m is the distance between the two coils and $v_{\text{s}}\approx 5000$ m/s is the sound velocity through copper at cryogenic temperatures.  It is apparent that the time for this to occur is on the order of hundreds of microseconds, an order of magnitude longer than the measured time delay.

The third alternative is not entirely different from the second, in that a heat transfer time scale will be calculated, however this will be a fully classical treatment using the diffusion equation,

\begin{equation}
\frac{\partial T}{\partial t}= \alpha \nabla^2 T = \alpha \frac{\partial ^2 T}{\partial z^2}
\end{equation}
where $T$ is the temperature of the sample, and $\alpha$ is the thermal diffusivity of copper. We consider only one spatial dimension $z$ in the Laplacian operator due to the wire geometry of the sample.  

At temperatures near $T=0$, the thermal diffusivity satisfies \cite{Jensen}

\begin{equation}
\alpha \approx 30 \text{ m}^2/\text{s}
\end{equation}

Using an (approximated) homogeneous initial condition across the length of the wire, and an inhomogeneous boundary condition at the high field SC coil (where we define $z=0$), we have

\begin{eqnarray}
T(z,t=0)&=&0 \\
T(z=0,t)&=&T_c \label{boundary-condition}
\end{eqnarray}
which is valid for the short time scales over which a flux jump occurs.

The Green's function for this model is found by spatially differentiating the heat kernel, and is given by
\begin{equation}
G(z,t)=\frac{z}{\sqrt{4 \pi \alpha t^3}}\exp\left(-\frac{z^2}{4\alpha t}\right)
\end{equation}
and the solution to the diffusion problem is the convolution of $G$ with the boundary condition (\ref{boundary-condition}) leading to
\begin{equation}
T(z,t)=T_c\int_0^{t}G(z,t-s)\ ds
\end{equation}
This integral can be evaluated analytically, and thus the temperature of the wire is given by
\begin{equation}
T(z,t) = T_c\text{ erfc}\left(\frac{z}{2\sqrt{\alpha t}}\right)\equiv T_c\left[1-\text{erf}\left(\frac{z}{2\sqrt{\alpha t}}\right)\right] \label{temp-solution}
\end{equation}
where the error function is defined as
\begin{equation}
\text{erf}(x)\equiv \frac{2}{\sqrt{\pi}} \int _0^x \exp(-x^2)\ dx
\end{equation}

Using the typical value of the critical temperature of $T_\text{c}\approx 10$ K for NbTi \cite{Blatt}, we wish to find the time at which $T = 7.5$ K, corresponding to where the persistent currents disappeared in the preliminary experiment depicted in Figure \ref{fig:figL0}.  For the full distance $z=l=3$ m, this time scale is on the order of seconds.  However, the excess wire between the two coils was itself made into loops, and if one ignores Kapitza resistance and assumes that the heat can travel from one section of the wire to another via a loose contact, a more conservative distance of $z=25$ cm should be used instead.  Note that it is not assumed that Kapitza resistance can be ignored when calculating the $L/R$ time constant, or in the phonon-interaction model.  For this smaller fixed distance of $z$, the time required for the temperature at the low field SC coil to reach 7.5 K is on the order of tens of ms, still multiple orders of magnitude above the measured time.

Thus, while many other potential theoretical and experimental research avenues exist on this topic, one can be guided by the likelihood that this process is fully quantum mechanical in nature, and the theoretical description outlined in Section II may be a valid starting point for future research.

\textbf{Acknowledgements}: We thank Kirk Wegter-McNelly for helpful discussions on theory.  SJM thanks Roman Mints for helpful discussions on theory and experiment.

\end{document}